\documentclass[a4paper,11pt]{article}
\usepackage{pos}

\title{Towards (semi) exclusive cross section measurements and modelling}

\author[a,b]{Maria B. Barbaro}

\affiliation[a]{Dipartimento di Fisica dell'Universit\`a di Torino and INFN - Sezione di Torino,\\
  Via Pietro Giuria 1, 10125 Turin, Italy}

\affiliation[b]{ Universit\'e Paris-Saclay, CNRS/IN2P3, IJCLab,\\
15 rue Georges Cl\'emenceau, 91405 Orsay cedex, France}

\emailAdd{barbaro@to.infn.it}

\abstract{Semi-inclusive neutrino-nucleus scattering, corresponding to the simultaneous detection of a lepton and one or more hadrons in the final state, is shown to be much more sensitive to nuclear effects than the inclusive process. The theoretical description of the semi-inclusive process is illustrated and some comparisons of relativistic plane wave impulse approximation (RPWIA) predictions with recent MicroBooNE, T2K and MINERvA data on argon and carbon are shown. The merit of new variables constructed to enhance particular nuclear effects is also illustrated and future developments going beyond the RPWIA approach are outlined.
}
\FullConference{%
  *** The 22nd International Workshop on Neutrinos from Accelerators (NuFact2021) ***\\
  *** 6–11 Sep 2021 ***\\
  *** Cagliari, Italy ***}

\begin{document}
\maketitle

\section{Introduction}

It is well recognised that nuclear physics plays an important role in the analysis of neutrino oscillation experiments exploring physics beyond the Standard Model. 
\!The extraction of the parameters characterising the mixing matrix U from the measured oscillation probability crucially depends on the knowledge of the neutrino energy, which must be reconstructed from the kinematics of the detected particles (leptons and hadrons). Since detectors are made of heavy nuclei, like carbon, oxygen and argon, accurate nuclear models are needed in the experimental analyses. The uncertainties related to nuclear effects represent today one of the largest sources of systematic error, which must be reduced to ensure the success of future experiments, notably DUNE \cite{DUNE:2020lwj} and T2HK \cite{Hyper-KamiokandeProto-:2015xww}.

The most general lepton-nucleus  scattering reaction is
\begin{equation}
l+A \rightarrow l'+B+X \,,
\end{equation}
where a lepton $l$ scatters off a nucleus $A$ in its ground state and a final state is produced, containing the scattered lepton $l'$, some hadronic system $X$ (for instance, one or more nucleons, pions or other mesons) and the residual nucleus $B$,  usually left in an excited state.
The theoretical description of this process represents a tough many-body problem, which cannot be solved exactly but only approximately, resorting to nuclear models. Each model involves approximations, the validity of which depends not only on the energy domain (ranging typically from the quasielastic regime dominant at T2K kinematics up to the deep-inelastic scattering one, which will be important at DUNE energies), but also on the type of experimental signal one wants to describe. According to the particles detected in the final state, the process is defined as  {\it inclusive} when only the outgoing lepton is measured,  {\it semi-inclusive} when $X$ is detected in coincidence with the lepton and {\it exclusive} when the complete final state is known, including the residual nucleus. Fully exclusive measurements can be performed in electron scattering experiments, where the beam energy is usually very well known, by choosing the kinematics of the observed final state and using energy and momentum conservation, but not in the neutrino case, where the beam is not monochromatic and the energy is distributed according to a more or less broad flux, according to the specific experiment.  

Most experimental and theoretical papers published up to now  (see \cite{NuSTEC:2017hzk} for a comprehensive review)  have dealt with the inclusive case, reporting cross sections as functions of the lepton variables only.  The ingredients and techniques of the available calculations are quite different: some  are based on various approaches to the Random Phase Approximation \cite{Martini:2009uj,Jachowicz:2002rr,Nieves:2011pp}, other ones are focused on a sophisticated description of the nuclear spectral function \cite{Benhar:2005dj}, on the Green's function Monte Carlo approach  trying to solve exactly the many-body problem \cite{Lovato:2017cux}, or on transport theory \cite{Leitner:2008ue}. Two models  put more emphasis on the relativistic aspect of the problem, which is crucial in the GeV regime typical of oscillation experiments, as well as on the validation with electron scattering results: the Relativistic Green Function model \cite{Meucci:2003cv} and the SuSAv2 model, based on the superscaling behaviour of electron scattering data and its description within relativistic mean field theory (see \cite{
Amaro:2021sec,Amaro:2019zos,Barbaro:2021psv} for recent reviews of this approach).
Nevertheless,  all these models provide rather similar predictions for inclusive  observables,  with a typical spread of 10-20\% \cite{Branca:2021vis}, while the present experimental precision does not allow for discriminating between them. 
On the contrary, more exclusive measurements can better constrain nuclear models and hence help to reduce the associated systematic error: when more particles are detected in the final state, the cross section becomes more sensitive to the details of the nuclear dynamics and more accurate models are needed to describe such processes.
Several recent measurements of semi-inclusive observables, involving the kinematics of both the final lepton and hadron(s), have been recently performed by the T2K, MINERvA and MicroBooNE collaborations \cite{
T2K:2018rnz,MINERvA:2018hba,MINERvA:2019ope,MicroBooNE1,MicroBooNE2}. A parallel effort is definitely needed on the theory side and this will likely be the object of more studies in future years. 

In what follows we shall summarise the basic formalism needed to attack the problem of semi-inclusive neutrino-nucleus scattering (Sect.\ref{sec:form}), show some results obtained within the Relativistic Plane Wave Approximation, as a first step of the extension of the  SuSAv2 model to the semi-inculsive case (Sect.\ref{sec:res}), and outline future developments and open problems (Sect.\ref{sec:concl}).

\section{Semi-inclusive neutrino-nucleus scattering}
\label{sec:form}

The full formalism for semi-inclusive neutrino nucleus reactions has been developed in Refs.~\cite{Moreno:2014kia,VanOrden:2019krz,Franco-Patino:2020ewa}. 
Let us consider the simplest semi-inclusive reaction, one-nucleon knock-out. 
The variables commonly used in electron scattering studies to describe this process are the missing energy and momentum
\begin{equation}
E_m  = \omega-T_N - T_{A-1} ,\ \ \ {\bf p}_m = {\bf q}-{\bf p_N} = {\bf p_{A-1}}\,,
\label{eq:empm}
\end{equation}
where $\omega$ and ${\bf q}$ are  the energy and momentum transferred to the nucleus, while  $T_N$ , ${\bf p_N} $, $T_{A-1}$ and ${\bf p_{A-1}}$ are  the kinetic energies and momenta of the ejected nucleon $N$ and of the recoiling nucleus $A$--1.
 In the Plane Wave Impulse Approximation (PWIA), where the probe interacts with one nucleon and the outgoing nucleon momentum is not modified by Final State Interactions (FSI), $E_m$ and $p_m$ are simply the energy and momentum of the bound nucleon.  Unlike the case of  monochromatic electron beams, where the  variables \eqref{eq:empm} can be fixed experimentally by choosing the hadron kinematics,  in neutrino experiments the missing energy and momentum are not directly measurable, although they are still good variables to study the process from the theoretical point of view.

The 6th-differential neutrino ($\chi=$+1) or antineutrino ($\chi=$-1) cross section with respect to the lepton and nucleon momenta and solid angles, $(k',\Omega')$ and $(p_N,\Omega_N)$, can be written as
\begin{equation}
\Big\langle \frac{d^6\sigma}{dk' d\Omega' dp_N d\Omega_N}\Big\rangle_\chi =  \int_0^\infty dk \frac{\Phi_\nu(k)}{k} K_0\,
 {\cal F}_\chi^2\, S(p_m,E_m)\,\theta(E_m-E_s) ,
\label{eq:d6s}
\end{equation}
where $k$ is the (anti)neutrino beam momentum, $\Phi_\nu$ the associated flux, $K_0$ a kinematic factor (see \cite{Franco-Patino:2020ewa} for the explicit expression) and $E_s = m_N+M_{A-1}-M_A$ the nucleon separation energy.

The function ${\cal F}_\chi^2$  arises from the contraction of the leptonic and nucleonic tensors and is the linear combination of 10 independent response functions,  each of them depending upon 5 variables, $R^K\equiv R^K(\omega,q,{\bf p_N})$:
\begin{eqnarray}
{\cal F}_\chi^2 &=& V_{CC} R^{CC} + 2 V_{CL} R^{CL} + V_{LL} R^{LL} + V_T R^T+V_{TT} R^{TT}+V_{TC}R^{TC}+V_{TL}R^{TL}
\nonumber\\
&+& \chi \left(V_{T'} R^{T'}+V_{TC'}R^{TC'}+V_{TL'}R^{TL'}\right) .
\end{eqnarray}
Note that this structure is  more complex than  the one corresponding to the inclusive case, which involves only 5 responses of the 2 variables $(\omega,q)$. The five missing responses --  $R^{TT,TC,TL,TC',TL'}$ --   are proportional to the $\cos(\phi_N)$ and $\cos(2\phi_N)$,  $\phi_N$ being the nucleon azimuthal angle, and they disappear when the integral over the full final hadron phase space is performed. This is an important point, showing explicitly that  the inclusive cross section can be obtained from the semi-inclusive one (providing, of course, that all the relevant processes are taken into account),  but not viceversa.

The one-hole spectral function $S$ appearing in Eq.~\eqref{eq:d6s} is defined as
\begin{equation}
S(p_m,E_m) = \langle A|a^\dagger(p_m)\ \delta(\hat H-E_0-E_m)\ a(p_m)|A\rangle ,
\end{equation}
being $\hat H$ the nuclear Hamiltonian, $E_0$ the ground state energy of the nucleus $|A\rangle$ and $a^\dagger, a$ the fermion creation and annihilation operators. The spectral function represents the joint probability of removing a nucleon with given momentum $p_m$ from the nucleus, leaving the residual system in a state characterised by $E_m$. Its integral over all missing energies yields the momentum distribution
\begin{equation}
n(p_m)=\int_0^\infty S(p_m,E_m) \, dE_m \,.
\label{eq:np}
\end{equation}

It should be mentioned that the factorised expression \eqref{eq:d6s} is not exact and is strictly valid in PWIA. However in most kinematics this is a very good approximation and a convenient representation of the physics, although un-factorised expressions are actually  used in some calculations.

\begin{figure}[!htbp] 
	\centering
	\includegraphics[width=0.28\textwidth]{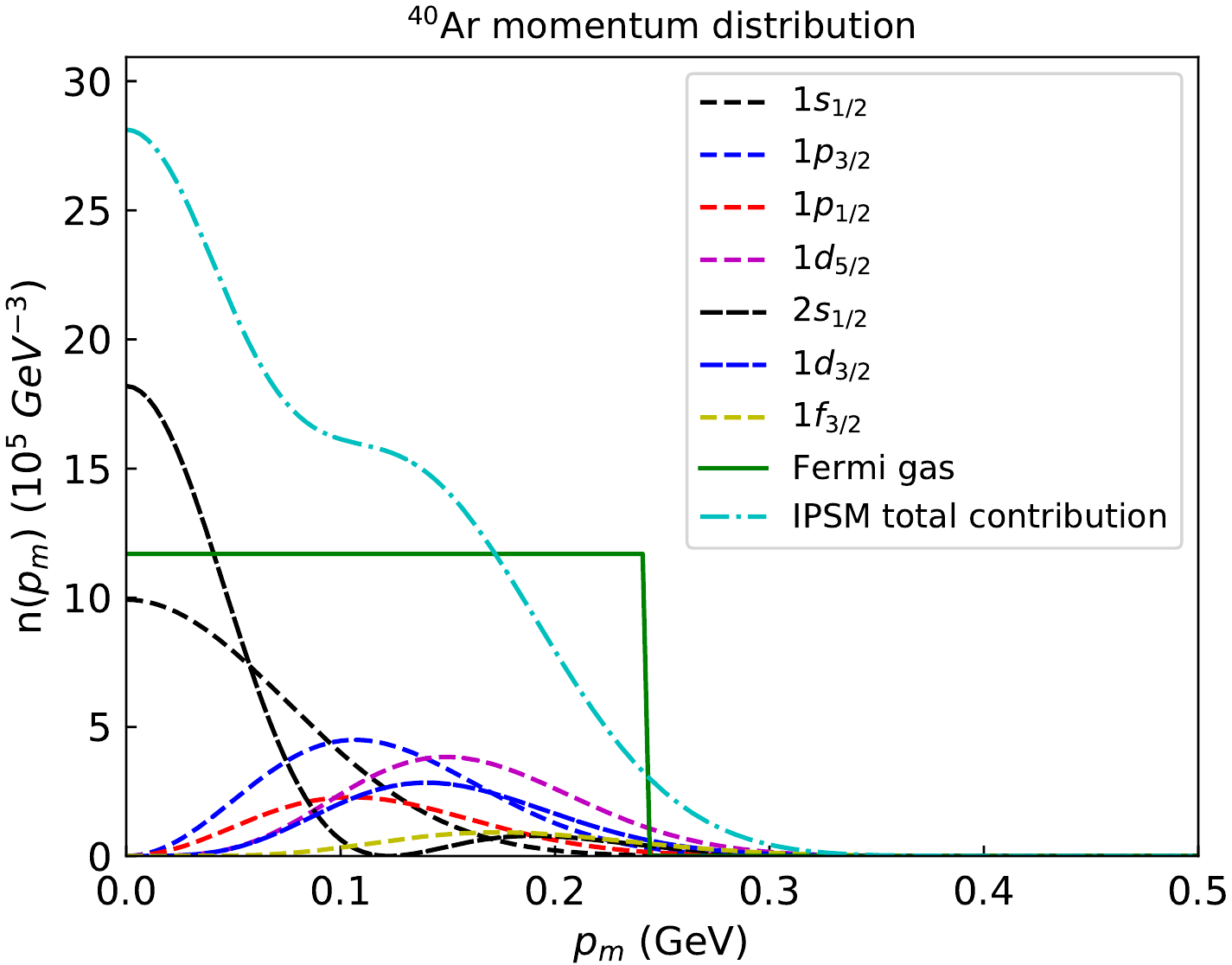} %
		\includegraphics[width=0.35\textwidth]{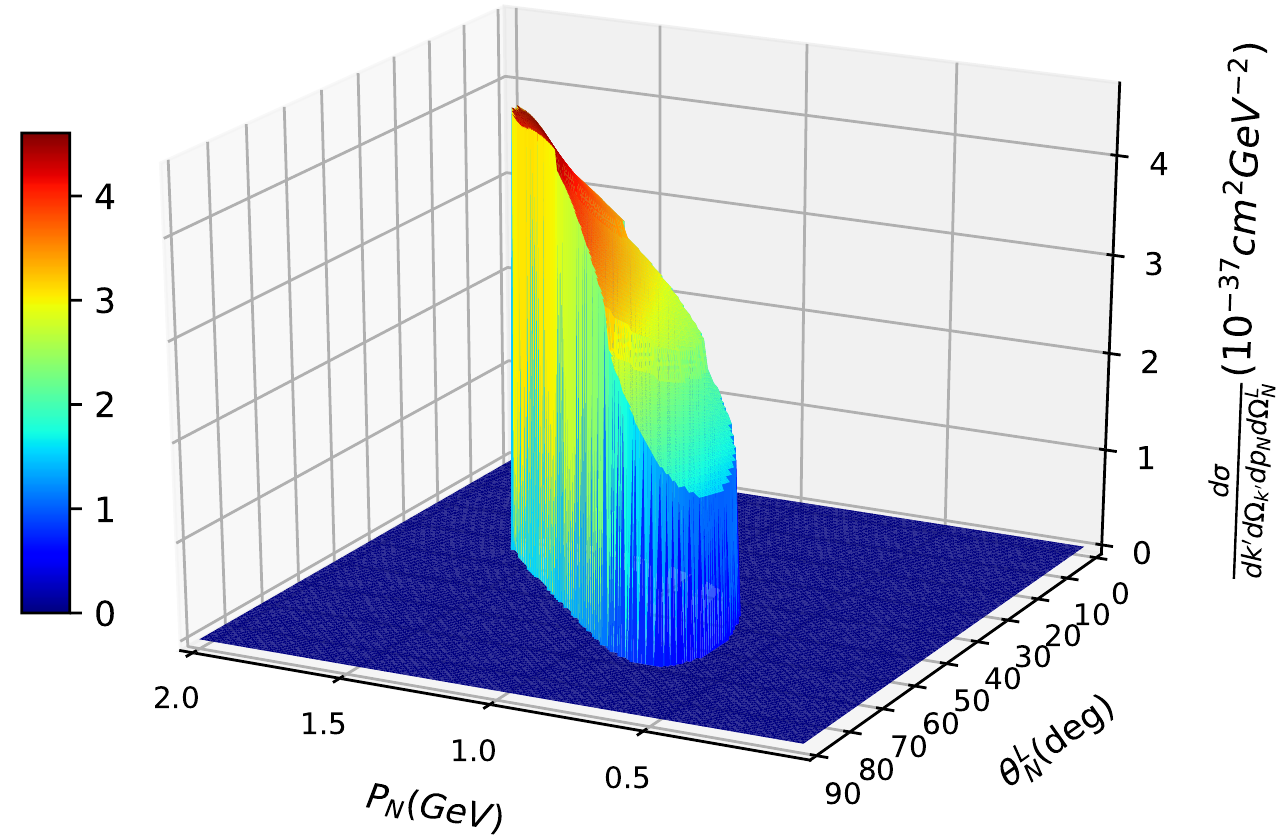} %
				\includegraphics[width=0.35\textwidth]{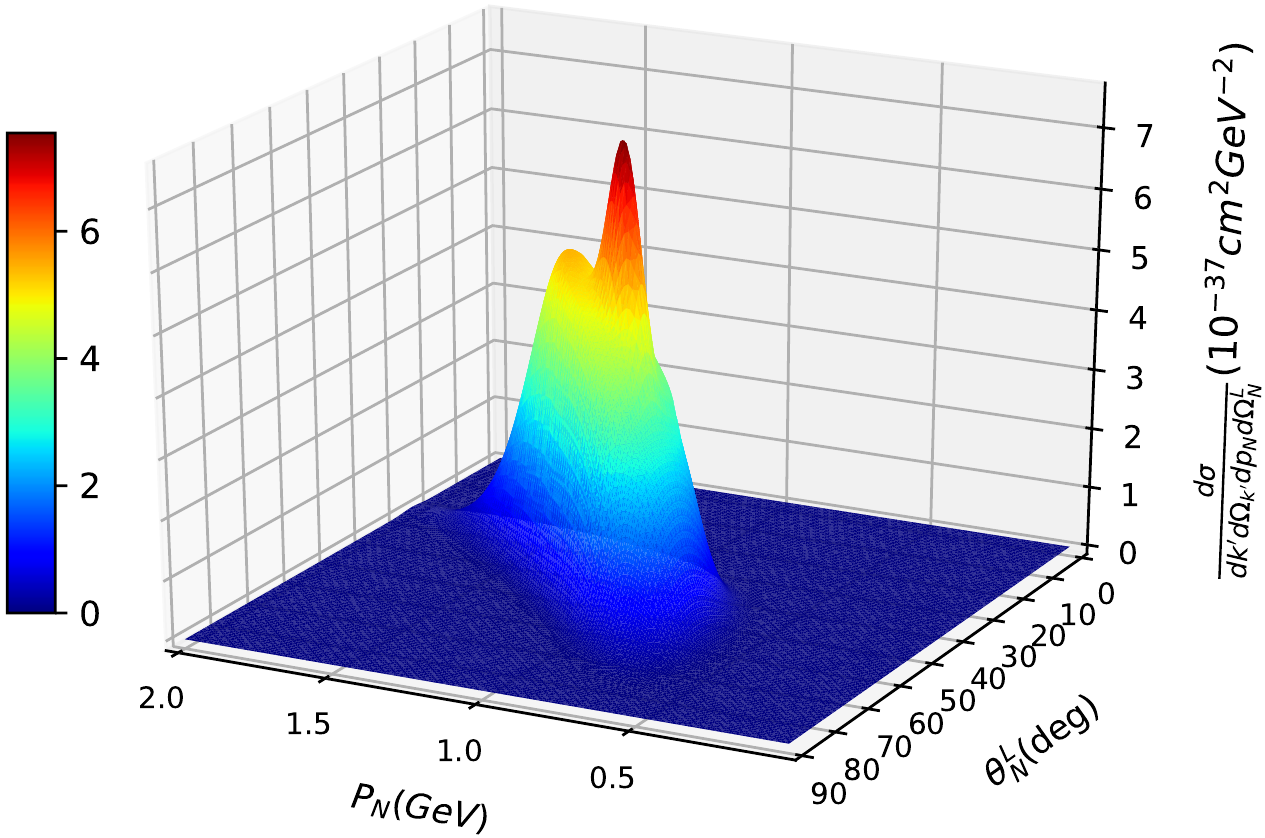} %
	\caption{\label{mom_dis_ar40} IPSM and RFG momentum distributions  for $^{40}$Ar (left panel) and the semi-inclusive $(\nu_\mu,\mu p)$ cross section evaluated in the RFG (middle) and IPSM (right) models, folded with the experimental DUNE flux, for $k'$=1.5 GeV/$c$, $\theta$=30$^\circ$ and $\phi_N$=$\pi$. Figure adapted from Ref.~\cite{Franco-Patino:2020ewa}.}
\end{figure} 
In Fig.\ref{mom_dis_ar40}  (left panel) we show the momentum distributions associated to two typical, very different, nuclear models: the Relativistic Fermi Gas (RFG) and the Independent Particle Shell Model (IPSM). In the RFG  the nucleons are treated as free relativistic fermions, correlated only by the Pauli principle; the associated spectral function
\begin{equation}
S_{\rm RFG}(p_m,E_m) = \theta(k_F-p_m)\,\delta\left(E_m-\sqrt{p_m^2+m_N^2}\right)\,,
\end{equation}
where $k_F$ is the Fermi momentum and $m_N$ the nucleon mass, leads, when inserted in \eqref{eq:np}, to a step-like momentum distribution. In the IPSM the spectral function is
\begin{equation}
S_{\rm IPSM}(p_m,E_m) = \sum_{nlj} (2j+1) n_{nlj}(p_m)\,\delta\left(E_m-E_{nlj}\right)\,,
\end{equation}
where the momentum distributions for each shell $njl$ are evaluated within the relativistic mean field model (RMF). The nucleus considered in Fig.\ref{mom_dis_ar40} is $^{40}$Ar and the Fermi momentum employed in the RFG calculation is  $k_F=$ 0.241 GeV/$c$.
In  the other two panels the 6th-differential semi-inclusive $(\nu_\mu,\mu p)$ cross section obtained in the RFG (middle panel) and IPSM (right panel) models and  folded with the DUNE flux  is represented versus the proton momentum $p_N$ and polar angle $\theta_N^L$ at $\phi_N$=$\pi$ and fixed muon kinematics ($k'$=1.5 GeV/$c$, $\theta$=30$^\circ$). These calculations do not include final state interactions of the outgoing nucleon with the residual system: the striking difference between the RFG and IPSM results clearly illustrates the extremely strong sensitivity of the semi-inclusive cross section to initial state physics.

\section{Results}
\label{sec:res}

In this Section we present a few representative comparisons of the RPWIA predictions with semi-inclusive neutrino scattering data. More results can be found in Refs.~\cite{Franco-Patino:2021yhd}  and \cite{JuanmaProc}.
\begin{figure*}[!htb]
	\centering
	\includegraphics[width=\textwidth]{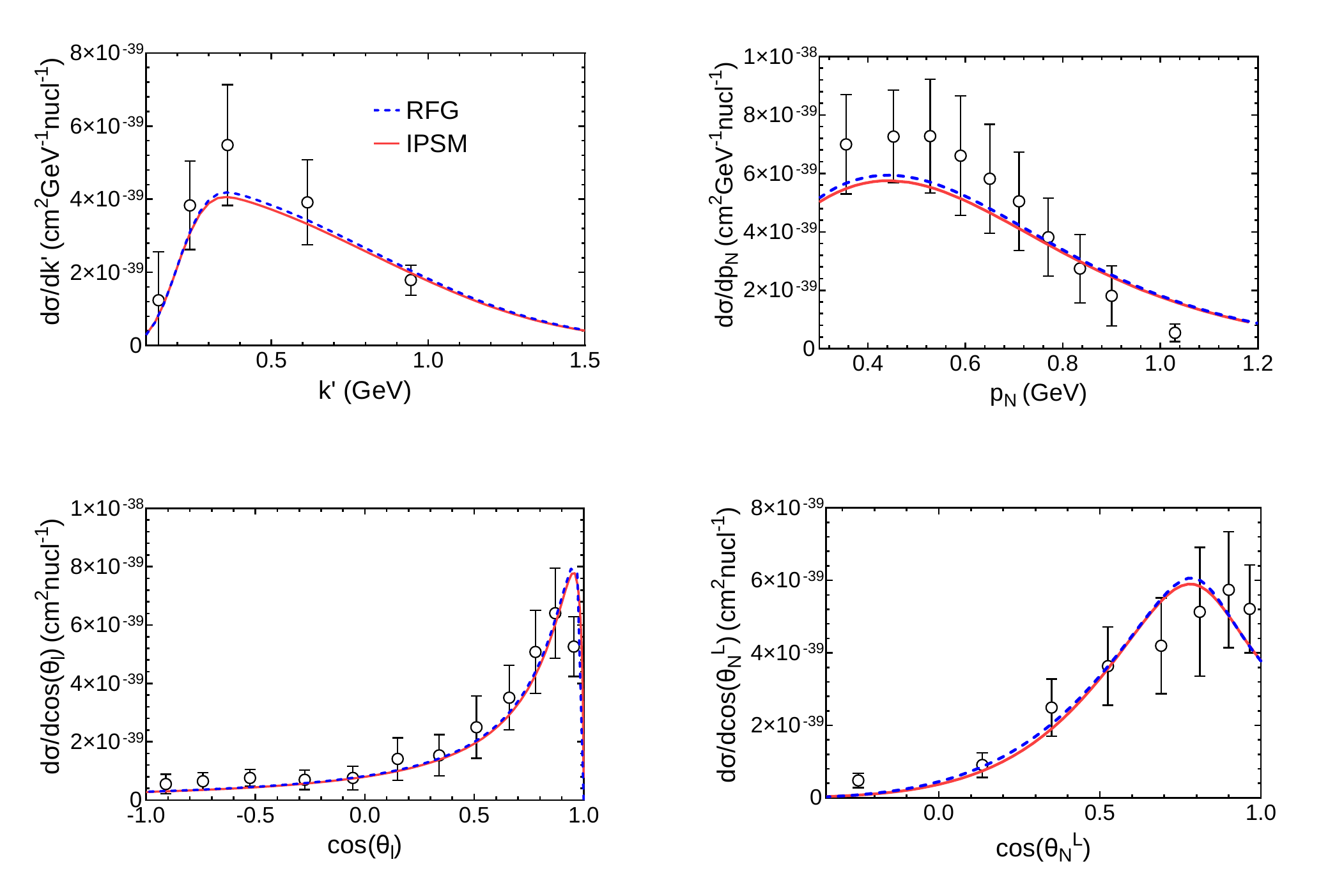}
	\caption{\label{fig:microboone A}The RPWIA semi-inclusive CC0$\pi$ single differential $\nu_\mu-^{40}$Ar cross sections displayed as functions of final muon and proton kinematics and compared with MicroBooNE data \cite{MicroBooNE1}, with the phase-space restrictions $k'$>0.1 GeV/$c$ and  0.3<$p_N$<1.2 GeV/$c$. Figure from Ref.~\cite{Franco-Patino:2021yhd}.}
\end{figure*}

In Fig.\ref{fig:microboone A} we show single differential charged-current $\nu_\mu-^{40}$Ar cross sections  with protons and no pions in the final state (CC0$\pi$), plotted versus  the muon (left panels) and proton (right panels) momenta and scattering angles. The curves corresponding to the two nuclear models above illustrated, RFG and IPSM, are compared with the MicroBooNE data \cite{MicroBooNE1}, with the phase-space restrictions $k'$>0.1 GeV/$c$ and  0.3<$p_N$<1.2 GeV/$c$. 

It is worth pointing out that the RPWIA calculations reproduce quite well the shape and magnitude of data, despite the fact that  two-particle-two-hole (2p2h) final  states, included in the data \cite{MicroBooNE1} but absent in the present calculation, are expected to significantly increase the cross section. On the other hand  the inclusion of FSI  yields in general lower cross sections, so that these two missing ingredients may lead to a  better agreement with data. The full calculation is in progress~\cite{inprep}.
It is also interesting to observe that the striking difference between the predictions of the RFG and IPSM, shown in Fig.\ref{mom_dis_ar40} for the coincidence cross section, is to a large degree washed out when the lepton or proton  variables are integrated over.

\begin{figure*}[!htb]
	\centering
	\includegraphics[width=0.35\textwidth]{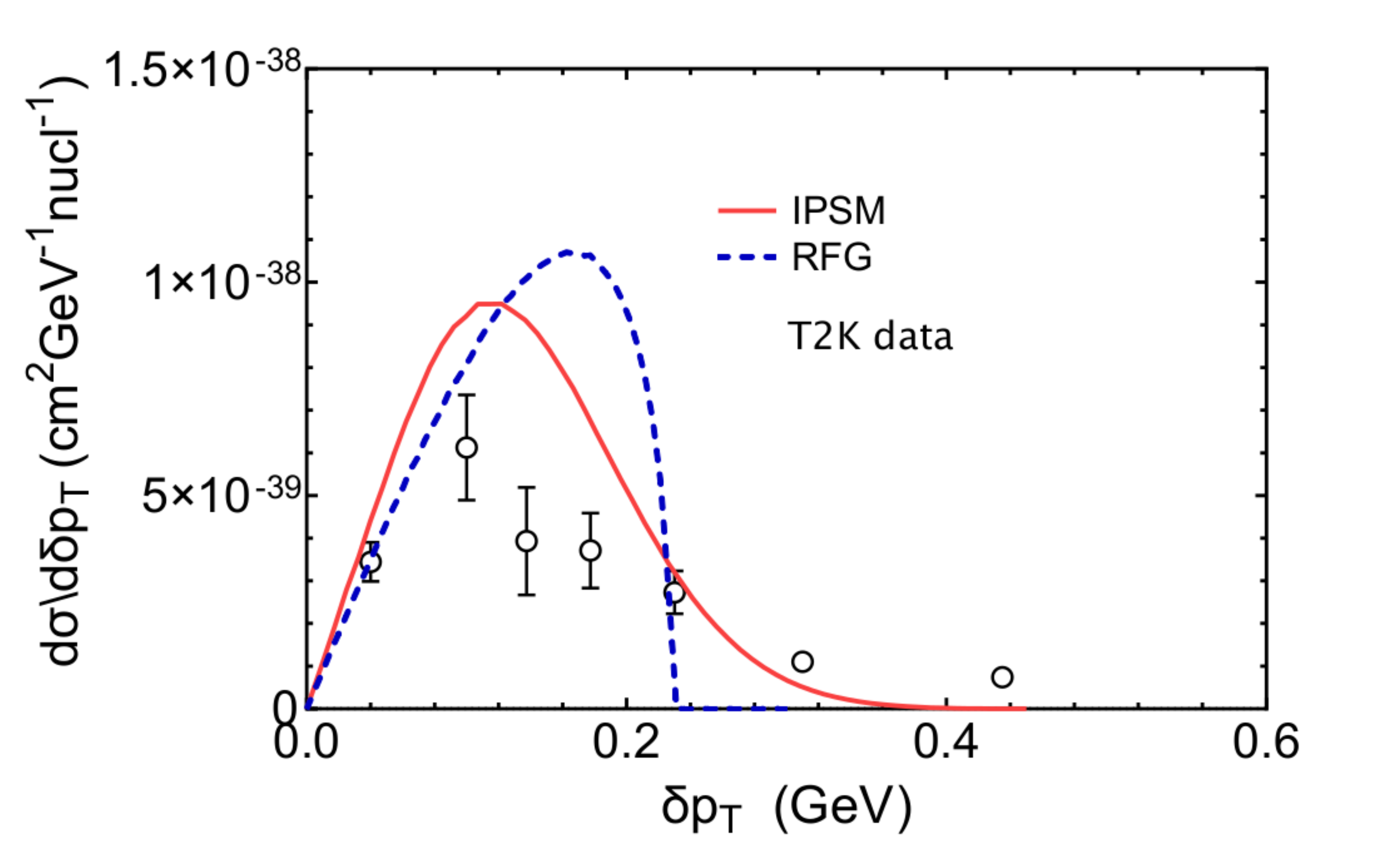}%
	\includegraphics[width=0.35\textwidth]{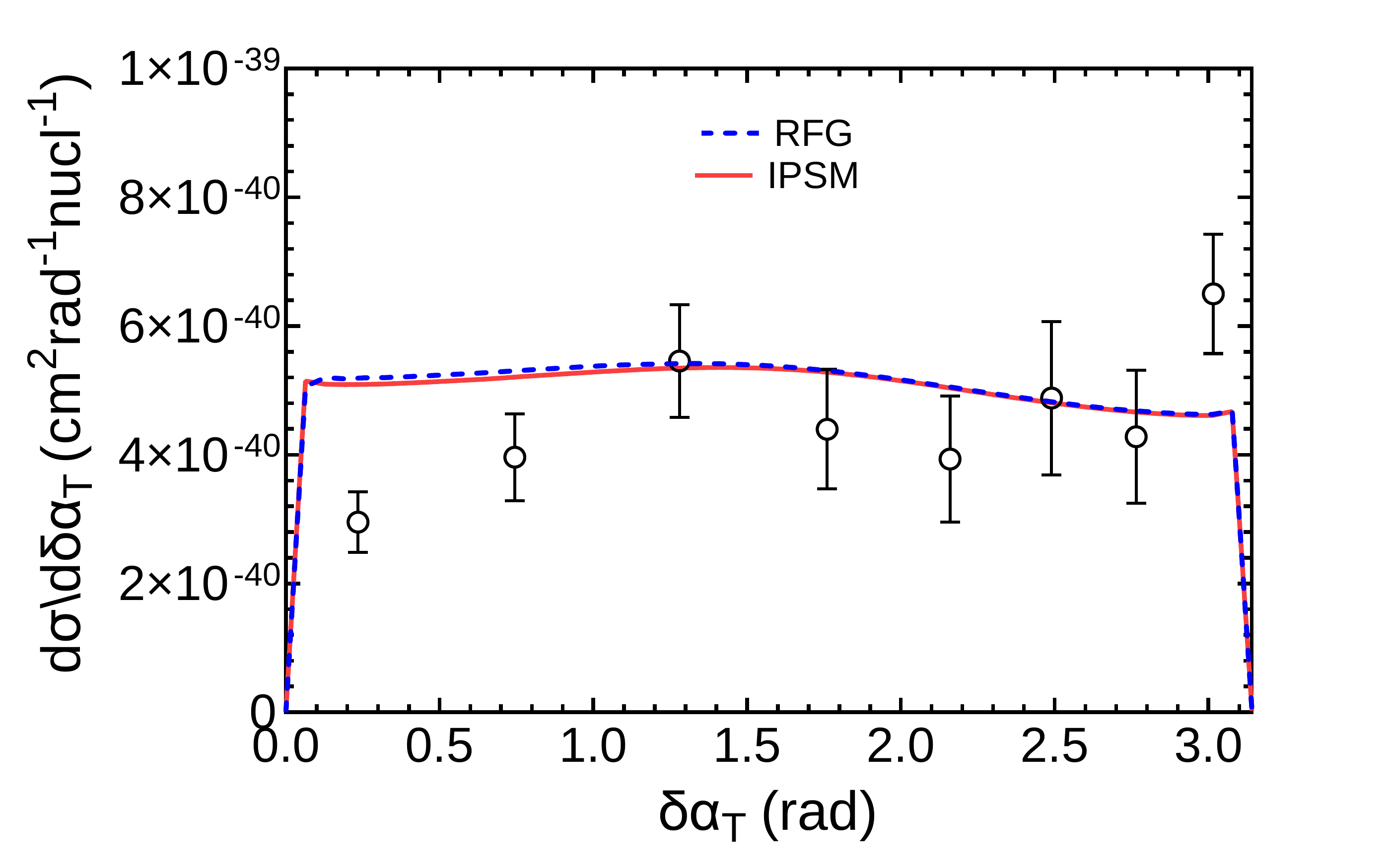}%
	\includegraphics[width=0.35\textwidth]{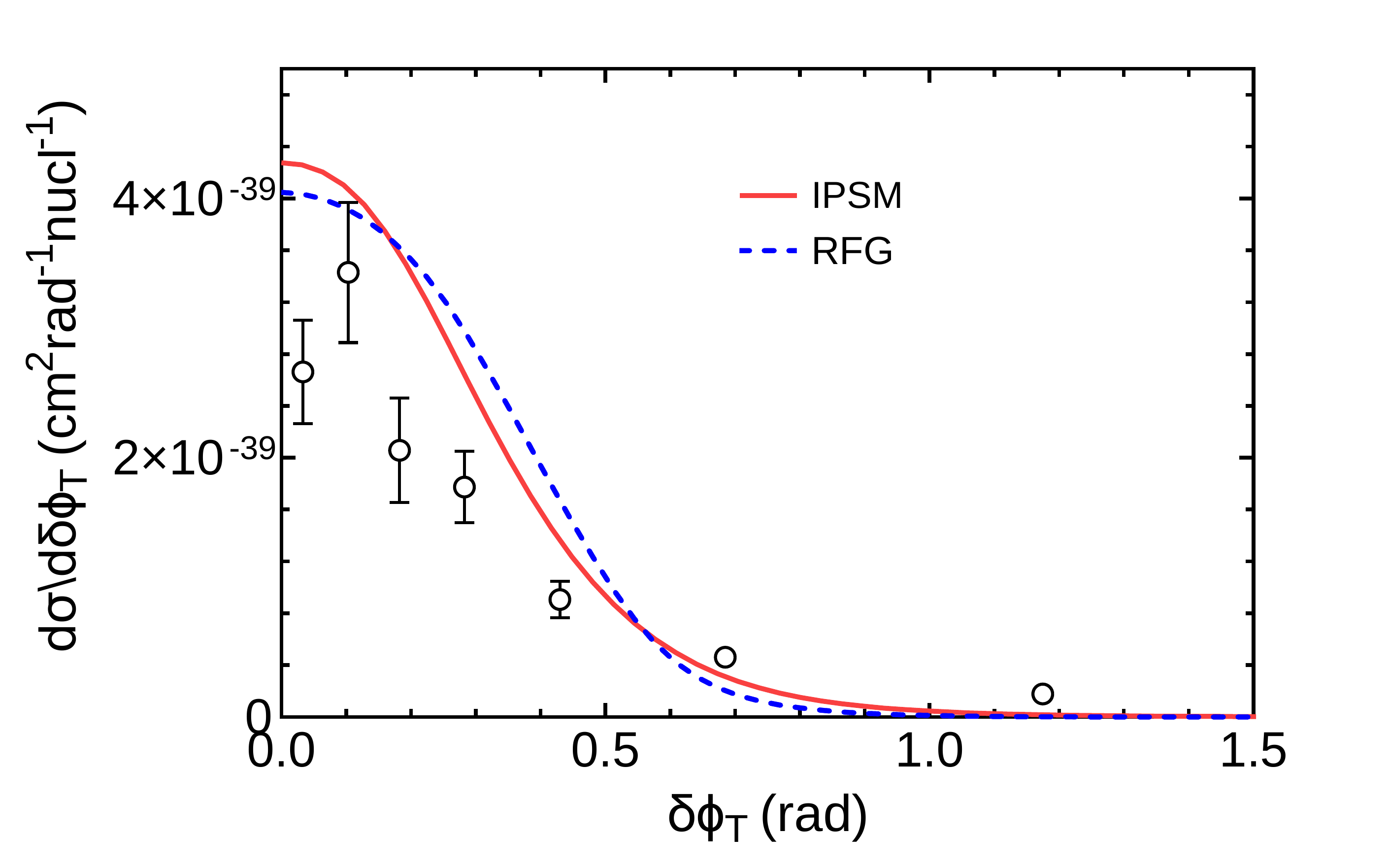}\\
	\includegraphics[width=0.35\textwidth]{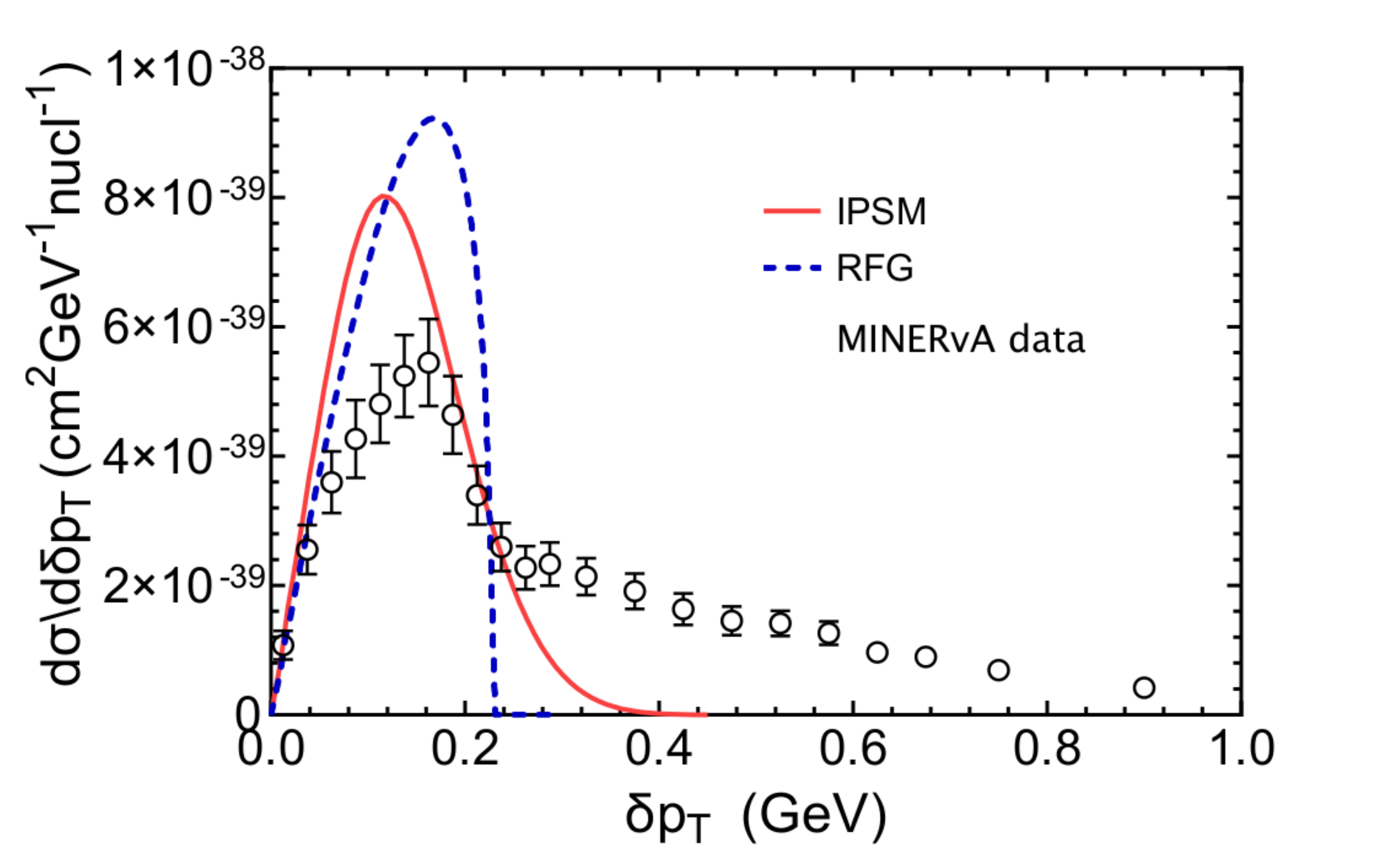}%
	\includegraphics[width=0.355\textwidth]{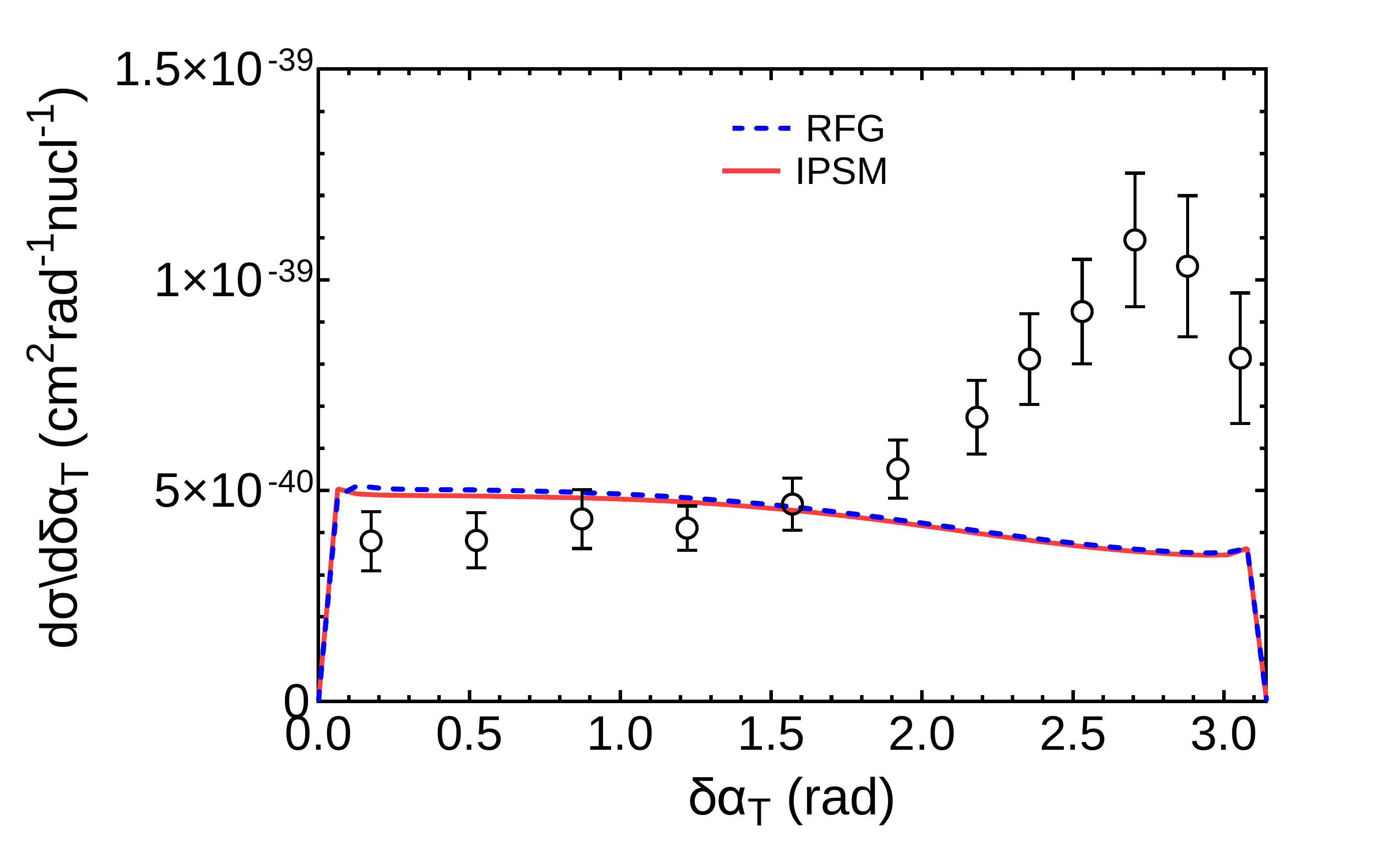}%
	\includegraphics[width=0.35\textwidth]{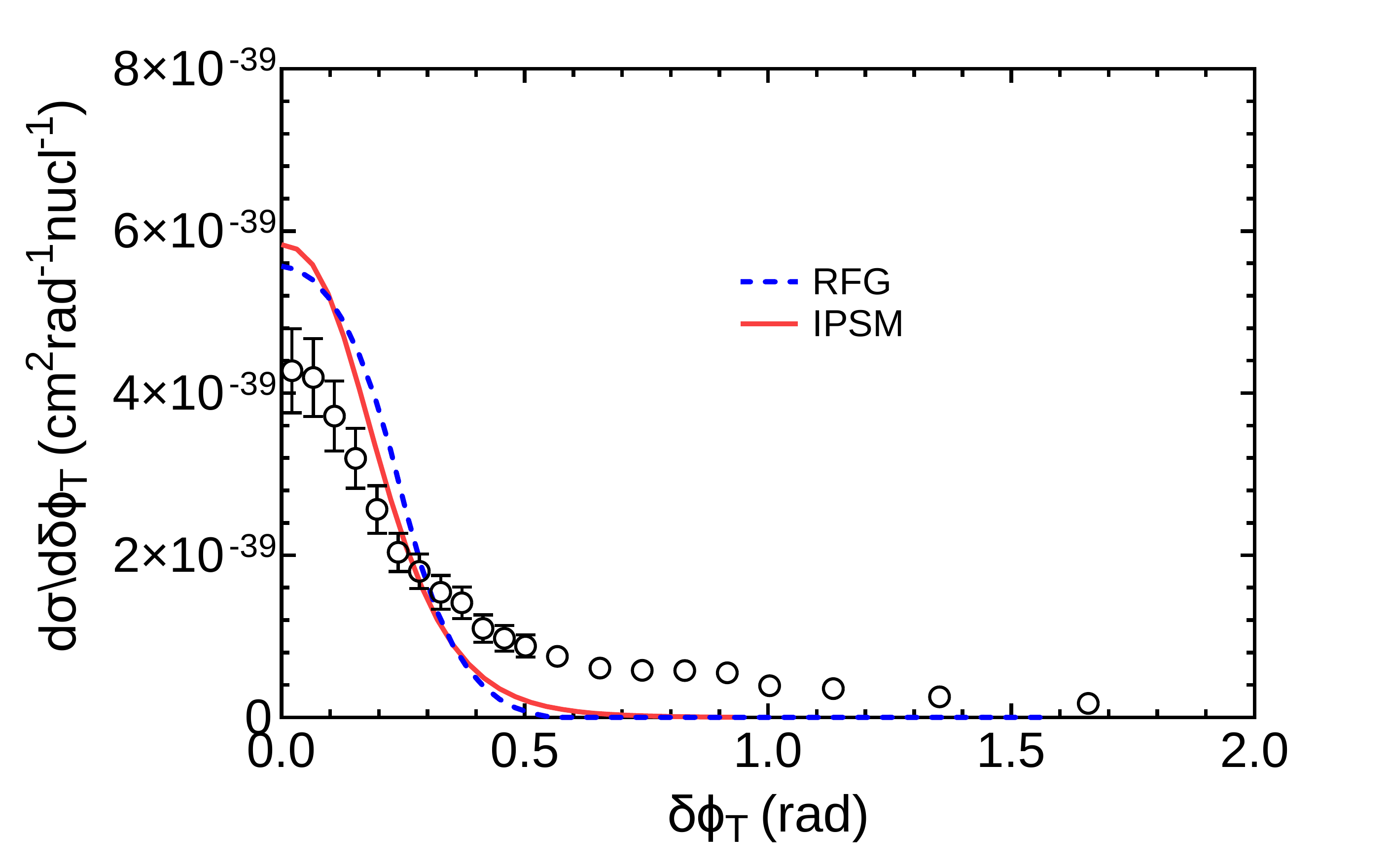}	
		\caption{\label{fig:STKI}The RPWIA semi-inclusive CC0$\pi$ single differential $\nu_\mu-^{12}$C cross sections displayed as functions of the variables \eqref{eq:STKI}  are compared with T2K (upper panels) and MINERvA (lower panels) data from Refs.~\cite{T2K:2018rnz} and \cite{MINERvA:2019ope}, respectively. Figure adapted from Ref.~\cite{Franco-Patino:2021yhd}.}
\end{figure*}

An enhanced sensitivity to nuclear effects can be achieved by introducing the single transverse kinematic imbalances (STKI) between the outgoing lepton and proton momenta in the plane transverse (T) to the neutrino beam. These are defined as~\cite{MINERvA:2018hba} 
\begin{equation}
{\bf \delta p_T} \equiv {\bf  k'_T}+{\bf  p_{NT}}\,,\ \ \ 
\delta\alpha_{\rm T} \equiv \arccos (-{\bf \hat k'_T}\cdot{\bf \hat \delta p_{T}})\,,\ \ \
\delta\phi_{\rm T} \equiv \arccos (-{\bf \hat k'_T}\cdot{\bf \hat p_{NT}})\,.
\label{eq:STKI}
\end{equation}
On a free nucleon at rest, the cross sections with respect to these variables would be either flat (for $\delta\alpha_{\rm T}$) or sharply peaked (for the other two variables), so that any deviation from this behaviour can be considered as a "measurement" of nuclear effects. This is confirmed by the results shown  in Fig.\ref{fig:STKI}, where the RPWIA predictions are compared  with T2K and MINERvA data for $\nu_\mu-^{12}$C scattering. Indeed the theoretical results deviate from the data more than those relative to the standard proton variables shown in Fig.\ref{fig:microboone A}. The cross section $d\sigma/d\delta p_{\rm T}$, which is closely related to the nucleon momentum distribution, is the only one sensitive to initial state physics. Moreover, as expected, the disagreement with the data, which should be ascribed to the missing  FSI and 2p2h effects, is stronger  for MINERvA, which operates at $\langle E_\nu\rangle\sim$3 GeV, than for T2K, which involves lower neutrino energies, $\langle E_\nu\rangle\sim$0.6 GeV.

\section{Conclusions and perspectives}
\label{sec:concl}

A new exciting season of experimental and theoretical activity has started in the neutrino scattering community, aimed at better constraining the nuclear models used in event generators through the measurement and modelling  of semi-inclusive observables.

While on the experimental  side many such data have been recently published and more will soon appear, theoretical papers on this subject are still rare.
The description of semi-inclusive processes requires a more complex formalism than the one needed in the inclusive case and involves all the components of the nuclear tensor, including those which do not contribute to the inclusive response. Furthermore,  a higher degree of sophistication  is required in the description of the nuclear dynamics: some models might give reasonable agreement with inclusive data but badly fail to reproduce observables which involve not only the lepton but also the hadron kinematics.
As a consequence, most nuclear models presently implemented in Monte Carlo generators have to be revised and improved.
A paradigmatic example is represented by the relativistic Fermi gas, which could be considered as an acceptable zero-th order approximation for the inclusive case, but is a completely inadequate framework to describe the semi-inclusive reaction.

As for the inclusive case, past studies of the electron scattering $(e,e'p)$ reaction represent a valuable benchmark, which must be extended to the neutrino case taking into account both the presence of the axial current and the different experimental setup, a more challenging configuration due to the flux average.

As a first step towards a complete description of semi-inclusive reactions, the RPWIA approach has been used to describe the $(\nu_\mu,\mu p)$ reaction. In this approach the initial state is treated using an independent particle shell model based on relativistic mean field theory, while the final proton state is approximated by a plane wave. Although the RPWIA calculation is obviously incomplete, it is useful in order to isolate the sensitivity of the various observables to the details of the nuclear initial state.
In spite of its simplicity, this model is able to reasonably describe single differential cross sections versus the proton variables. However, it fails to match the observables measured as functions of combined lepton/hadron variables, the single transverse kinematic imbalances, which are particularly sensitive to nuclear effects beyond  the plane-wave approach (final state interactions - FSI) and beyond the impulse approximation (many-nucleon excitations induced by  two-body currents - 2p2h). The STKI variables carry sensitive information on the nuclear dynamics, which deserves further investigation.

The implementation of FSI in the model is not a trivial task, since the description of different experimental signals may require  the use of different approaches~\cite{Gonzalez-Jimenez:2021ohu}. A common way of incorporating FSI, widely used is electron scattering studies, consists in the use of complex optical potentials (OP), fitted to elastic proton-nucleus data  on various targets in a certain energy range. The imaginary part of the OP accounts for the flux lost in unobserved inelastic open channels. This approach has been extensively and successfully applied in the past to the exclusive $(e,e'p)$ reaction, where there is certainty that the final state contains one and only one proton. In the case of neutrino experiments the situation is further complicated by the fact that completely exclusive measurements cannot be performed due to the broad neutrino flux.
On the other hand the full complex OP  cannot be used in the inclusive case, where all the final states contribute to the cross section. 
For this reason only the real part of the potential is sometimes employed to describe inclusive reactions: although this recipe yields a reasonably good agreement with the data, it remains rather unsatisfactory from the theoretical point of view since it treats inconsistently the initial and final states. A more consistent method is adopted in the RMF approach, in which the final nucleon wave function is built as a scattering solution of the same relativistic Hamiltonian used to describe the initial bound state. This approach has the merit of respecting orthogonality between the initial and final states, but must be corrected in order to smooth out the RMF potentials, which are unrealistically strong at high energies, where the PWIA must be recovered.
The implementation of FSI in the RPWIA model is in progress~\cite{inprep} and will constitute a step forward towards the reduction of nuclear uncertainties in the analyses  of neutrino oscillation experiments.

\acknowledgments
This work was supported by the University of Turin under the project BARM-RILO-20-01 and by the Istituto Nazionale di Fisica Nucleare under the project
NucSys. The author thanks  Juan Manuel Franco-Pati\~no for his help in generating the plots.

\end{document}